\documentclass[journal]{IEEEtran}
\IEEEoverridecommandlockouts

\usepackage{amssymb, amsthm, amsmath, graphicx, cite}
\usepackage[caption=false,font=footnotesize]{subfig}
\usepackage[ruled]{algorithm2e}

\newtheorem{theorem}{Theorem}

\newtheorem{corollary}{Corollary}

\newcommand{\MSE}{{\sf MSE}}

\newcommand{\D}{{\sf D}}
\newcommand{\DP}{{\sf DP}}
\newcommand{\B}{{\sf B}}

\let\oldbrace\{
\def\{{\oldbrace\kern0.5pt}

\newcommand{\nn}{\nonumber}

\newcommand{\Cc}{\mathcal{C}}

\newcommand{\Nc}{\mathcal{N}}

\newcommand{\Rc}{\mathcal{R}}
\newcommand{\Sc}{\mathcal{S}}
\newcommand{\Tc}{\mathcal{T}}

\newcommand{\Xc}{\mathcal{X}}
\newcommand{\Yc}{\mathcal{Y}}

\newcommand{\Mh}{{\hat{M}}}

\newcommand{\mh}{{\hat{m}}}
\newcommand{\sh}{{\hat{s}}}

\newcommand{\Xt}{{\tilde{X}}}
\newcommand{\Yt}{{\tilde{Y}}}
\newcommand{\Zt}{{\tilde{Z}}}

\newcommand{\xt}{{\tilde{x}}}
\newcommand{\yt}{{\tilde{y}}}

\def\eps{\epsilon}

\DeclareMathOperator\E{\sf E}
\let\P\relax
\DeclareMathOperator\P{\sf P}

\begin{document}
\title{Integrated Communication and Bayesian Estimation of Fixed Channel States}

\author{Daewon Seo \\
	\thanks{ D.~Seo is with the Department of Electrical Engineering and Computer Science, Daegu Gyeongbuk Institute of Science and Technology (DGIST), Daegu 42988, South Korea (e-mail: dwseo@dgist.ac.kr).}
}

\maketitle

\allowdisplaybreaks

\begin{abstract}
	This work studies an information-theoretic performance limit of an integrated sensing and communication (ISAC) system where the goal of sensing is to estimate a random continuous state. Considering the mean-squared error (MSE) for estimation performance metric, the Bayesian Cram\'{e}r-Rao lower bound (BCRB) is widely used in literature as a proxy of the MSE; however, the BCRB is not generally tight even asymptotically except for restrictive distributions. Instead, we characterize the full tradeoff between information rate and the exact MSE using the asymptotically tight BCRB (ATBCRB) analysis, a recent variant of the BCRB. Our characterization is applicable for general channels as long as the regularity conditions are met, and the proof relies on constant composition codes and ATBCRB analysis with the codes. We also perform a numerical evaluation of the tradeoff in a variance estimation example, which commonly arises in spectrum sensing scenarios.
\end{abstract}
\begin{IEEEkeywords}
	Integrated sensing and communication, Fisher information, Bayesian Cram\'{e}r-Rao bound, asymptotically tight Bayesian Cram\'{e}r-Rao bound, constant composition codes
\end{IEEEkeywords}
\vspace{-0.1in}

\section{Introduction}
In next-generation wireless technologies like sixth-generation (6G) networks, integrated sensing and communication (ISAC) is expected to be a key enabler for convergent applications~\cite{Bourdoux2020_2}. Accurate parameter estimation, such as determining distance, velocity, and beamforming angles, is essential for reliable environmental sensing and communication~\cite{Liu--Masouros--Petropulu--Griffiths--Hanzo2020, Liu--Huang--Li--wan--Li--Han--Liu--Du--Tan--Lu--Shen--Colone--Chetty2022_2}. For example, in automotive communication systems, accurately estimating a vehicle's speed and position with a single data-bearing signal can significantly reduce hardware costs and required wireless resources.

As a single transmission signal should achieve two (perhaps conflicting) goals simultaneously, ISAC systems have two distinct performance metrics. For communication performance, Shannon's information rate is widely used. For unknown parameter estimation scenarios, it is common in literature to consider the mean-squared error (MSE) for a performance metric but use the Cram\'{e}r-Rao bound (CRB) as a performance proxy \cite{ChengLSHL2019}. This approach is valid since the CRB lower bounds the MSE and has several enjoyable properties, such as tractability and asymptotic tightness \cite{VanTrees1968}. For scenarios where the prior of the state is known, \cite{Xiong--Liu--Cui--Yuan--Han--Caire2023} and \cite{XuZ2024} use the Bayesian CRB (BCRB) as a performance proxy and analyze the tradeoff between the information rate and the BCRB. The BCRB is also employed in \cite{WangLLX2023}, where the primary performance metric, ``sensing estimation rate,'' is newly proposed based on rate-distortion theory, but the BCRB is adopted as a tractable proxy. Although the BCRB is a natural extension of the CRB in the Bayesian framework, it is not generally tight, even in asymptotic or high signal-to-noise ratio (SNR) regimes~\cite{Aharon--Tabrikian2024}. As a result, the performance bound defined by the BCRB could be significantly discrepant from the actual MSE performance.

Unlike these existing works, this study directly characterizes the tradeoff between Shannon's information rate and the \textit{exact} MSE by using the recently developed asymptotically tight BCRB (ATBCRB) as a lower bound on MSE \cite{Aharon--Tabrikian2024}. The significance of this study lies in the right performance metric for estimation and the explicit characterization of the optimal tradeoff. The key contributions and comparisons with existing works are summarized as follows.
\begin{itemize}
	\item We consider an ISAC problem where the objective of sensing is to estimate a random continuous state. We assume that a state is randomly drawn but fixed throughout the transmission as in \cite{TengYWJ2022, AnLNY2023}. Our model is information-theoretically general; we consider general communication and sensing channels under the regularity conditions and not limited to Gaussian-related ones.
	
	\item To the best of our knowledge, this is the first information-theoretic work that characterizes the tradeoff between the information rate and the \textit{exact} MSE, unlike existing literature that approaches MSE indirectly through the CRB \cite{ChengLSHL2019} or the BCRB \cite{Xiong--Liu--Cui--Yuan--Han--Caire2023, XuZ2024}. In particular, the BCRB is generally not tight unless the posterior distribution is Gaussian \cite{Aharon--Tabrikian2024}. Hence, for example, a code that achieves the proposed rate-BCRB tradeoff in \cite{Xiong--Liu--Cui--Yuan--Han--Caire2023} does not accurately capture the corresponding true rate-MSE tradeoff for general models, while ours provably achieves the true rate-MSE tradeoff.
	
	\item The expression for our rate-MSE tradeoff is as simple as that by the BCRB. From a technical perspective, our characterization is information-theoretically tight, meaning that the achievability and converse bounds are met. The achievability is established through constant composition codes (CCCs) \cite{Csiszar--Korner2011} and the use of maximum likelihood (ML) or maximum a posteriori (MAP) estimator to achieve the ATBCRB. The converse is derived by the fact that any codebook of length $n$ can be approximated by a CCC, and the estimation lower bound is provided by the ATBCRB for CCC.
	
	\item We illustrate with a spectrum sensing example that our tradeoff region defined in Thm.~\ref{thm:ATBCRB_ISAC} is significantly smaller than its rate-BCRB counterpart. Since ours is the optimal region, and a smaller region corresponds to a larger MSE at a fixed information rate, this suggests that the rate-BCRB proxy is substantially suboptimal compared to the true rate-MSE tradeoff for general models \cite{Xiong--Liu--Cui--Yuan--Han--Caire2023, XuZ2024}.
\end{itemize}

\section{Formulation \& Preliminary}
\subsection{Formulation}
\begin{figure}[t]
	\begin{center}
		\includegraphics[width=2.4in]{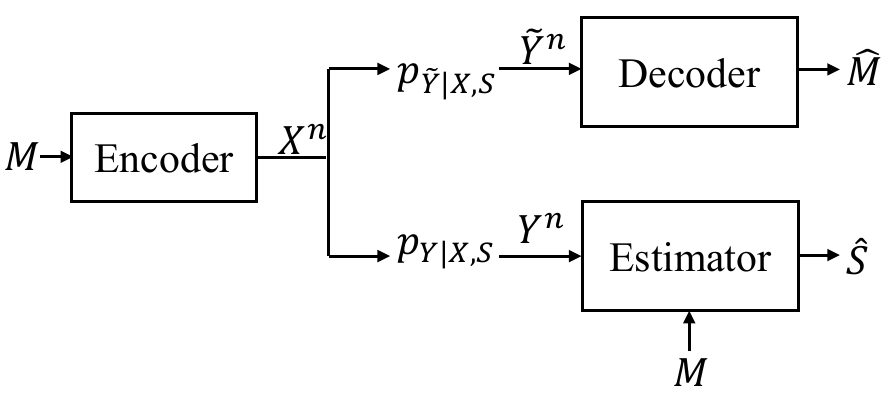}
	\end{center}
	\vspace{-0.2in}
	\caption{Problem model.}
	\vspace{-0.1in}
	\label{fig:model}
\end{figure}

Consider a bistatic communication and sensing model depicted in Fig.~\ref{fig:model}. This setting is widely used in literature \cite{Ahmadipour--Kobayashi--Wigger--Caire2022, Xiong--Liu--Cui--Yuan--Han--Caire2023, Chang--Wang--Erdogan--Bloch--2023}, where the encoder and estimator could be two base stations that are connected via a wireline backhaul link sharing the transmission signal. The presence of the message side information removes the uncertainty of the transmission signal, allowing the estimator to focus solely on the state estimation. It also can be thought of as monostatic ISAC without feedback codes (e.g., open-loop coding \cite{Chang--Wang--Erdogan--Bloch--2023}) if the encoder and the estimator are equipped on a single device. In this case, $p_{Y|X,S}$ is the backscatter channel.

Specifically, for a given codeword $x^n$ of length $n$, the channel distribution follows 
\begin{align*}
	p(s, \yt^n, y^n|x^n) = p(s) \prod_{i=1}^n p(\yt_i|x_i, s)p(y_i|x_i, s)
\end{align*}
where  $s\in\Sc \subset \mathbb{R}$ is a state randomly drawn from a known distribution $p(s)$, $\yt^n \in \tilde{\Yc}^n$ is an observation sequence at the receiver for communication, and $y^n \in \Yc^n$ is an observation sequence at the estimator for sensing. Two channels $p(\yt|x,s):=p_{\Yt|X,S}(\yt|x,s), p(y|x,s):=p_{Y|X,S}(y|x,s)$ are not necessarily identical. We assume that $S$ remains constant throughout the transmission. It captures physical changes in the channel that occur much more slowly than the transmission time scale, e.g., slow or block fading \cite{Chang--Wang--Erdogan--Bloch--2023, TengYWJ2022, AnLNY2023}.

The transmitter wishes to send a message $m \in [1:2^{nR}]$ to the communication receiver (decoder) while also allowing the estimator to estimate $S$. For the estimator, it is assumed that the message $m$ is available as side information. Formally, a $(2^{nR}, n)$ code of length $n$ for the joint communication and estimation problem consists of
\begin{itemize}
	\item a message set $m \in [1:2^{nR}]$, 
	\item an encoder that assigns a sequence $x^n(m) \in \Xc^n$ to each message $m\in[1:2^{nR}]$, 
	\item a decoder that assigns a message estimate $\mh(\yt^n)$ to each observation sequence $\yt^n$, and
	\item an estimator that assigns a state estimate $\sh(y^n | x^n(m))\in\Sc$ to each $y^n$ and $m$.
\end{itemize}
Note that $x^n$ is independent on $Y_i$, i.e., we consider open-loop coding \cite{Chang--Wang--Erdogan--Bloch--2023}. We denote the codebook by $\Cc^{(n)} = \{x^n(m), m\in[1:2^{nR}]\}$. It is assumed that transmitter and receiver alphabet spaces are finite, i.e., $|\Xc|, |\Yc|, |\tilde{\Yc}| < \infty$. However, we assume that $\Sc$ is continuous and $I(X;\Yt)$ for state $s$ is continuous in $s$ and finite. It enables us to approximate the mutual information arbitrarily close to the true value by discretizing $S$. Furthermore, we impose the standard regularity conditions on $p(s)$ and $p(y|x,s)$ so that the Cram\'{e}r-Rao bound and its variants are well established~\cite{VanTrees1968}.

The performance metrics are defined as follows. The probability of error for communication is defined as $P_e^{(n)} = \P(M\neq \Mh)$, i.e., the average probability of error. We measure the estimation performance of $\sh(y^n|x^n)$ by average mean-squared error (MSE): For a given codeword $x^n$,
\begin{align*}
	\MSE(\sh|x^n) = \E [ (S-\sh(Y^n|x^n))^2 ],
\end{align*}
where the expectation is over $S$ and $Y^n$. Since the state is estimated from $n$ noisy observations (and given information of $x^n$), it is expected that the MSE vanishes at $\Theta(n^{-1})$, cf.~estimation under i.i.d.~additive Gaussian noise~\cite{VanTrees1968}. To this end, we say that a rate-MSE decay pair $(R, \alpha)$ is achievable if there exists a sequence of $(2^{nR}, n)$ codes such that $\lim_{n\to\infty} P^{(n)}_e = 0$ and
\begin{align*}
	\limsup_{n \to \infty} \max_{x^n \in \Cc^{(n)}} n \MSE(\sh | x^n) \le \alpha
\end{align*}
Then, the rate-MSE decay region $\Rc$ is the closure of the set of all achievable pairs $(R, \alpha)$, that is 
\begin{align*}
	\Rc := \text{cl}\{(R,\alpha) \in \mathbb{R}_+^2: (R,\alpha) \text{ is achievable} \}.
\end{align*}
Therefore, our goal is to find the exact characterization of $\Rc$, i.e., find the optimal tradeoff between the data rate $R$ and the (normalized) MSE $\alpha$. It should be remarked that $\Rc$ directly addresses the MSE, rather than considering a (B)CRB-related term as a proxy as in \cite{Xiong--Liu--Cui--Yuan--Han--Caire2023, XuZ2024}.

\subsection{Preliminary}
This subsection provides a brief introduction to several key MSE lower bounds commonly used in literature.

First, when $p(s)$ is unavailable, i.e., in the non-Bayesian framework, the Cram\'{e}r-Rao bound provides a lower bound on the MSE as follows \cite{VanTrees1968}: For any \textit{unbiased} estimator $\sh$, under the regularity conditions,
\begin{align*}
	\E[(s-\sh(Y^n))^2] \ge \E^{-1} \left[ \left( \frac{\partial}{\partial s} \log p(Y^n|s) \right)^2 \right] =: J_{\D}^{-1}(s),
\end{align*}
where $J_{\D}(s)$ represents the Fisher information for (nonrandom) $s$. This bound is widely accepted as it is asymptotically tight. Here, ``asymptotically'' implies that the ``information'' in observations about $s$ grows without bound---such as when the signal-to-noise ratio (SNR) grows or $n \to \infty$, e.g., through coherent combining.

Second, when $p(s)$ is available, i.e., in the Bayesian framework, the Bayesian Cram\'{e}r-Rao bound (BCRB) is given as follows \cite{VanTrees1968}: For \textit{any} estimator $\sh$,
\begin{align}
	\E[(S-\sh(Y^n))^2] \ge \left( \E [ J_{\D}(S) + L_{\P} (S)] \right)^{-1} =: (J_B)^{-1}, \label{eq:BCRB}
\end{align}
where $L_{\P} (s) := (\frac{\partial}{\partial s} \log p(s))^2$ and $J_\B$ is referred to as the Bayesian Fisher information or the total Fisher information. While this bound provides a valid lower bound on MSE, it is generally not tight even in the asymptotic regime. Specifically, it is tight only when the posterior distribution is Gaussian, or the bound is asymptotically achievable only when $J_{\D}(s)$ is independent of $s$ \cite{Aharon--Tabrikian2024}.

Numerous efforts have been made to tighten the gap in the Bayesian framework, such as the Ziv-Zakai family \cite{ZivZ1969}, the Weiss-Weinstein family \cite{WeissW1985}, and their variants. However, these bounds require some additional efforts to obtain explicit characterization, which limits their practical insights. Recently, Aharon and Tabrikian proposed an asymptotically tight variant of the Bayesian Cram\'{e}r-Rao bound (ATBCRB)~\cite{Aharon--Tabrikian2024}. This variant offers a simple analytic expression that is asymptotically attainable by the ML or MAP estimation. Specifically, letting $J_{\DP}(s) := J_{\D}(s) + L_{\P}(s)$, the bound is as follows \cite[Thm.~5]{Aharon--Tabrikian2024}:
\begin{align}
	&\E[(S-\sh(Y^n))^2] \nn \\
	&\ge \frac{ \E^2[ J_{\DP}^{-1}(S)] }{ \E[ J_{\DP}^{-1}(S)] + \E\left[ \left( \frac{\partial }{\partial S} J_{\DP}^{-1}(S) \right)^2 \right] + \E\left[ \frac{\partial^2 }{\partial S^2} J_{\DP}^{-2}(S) \right] }, \label{eq:ATBCRB}
\end{align}
which is asymptotically tight by using the ML or MAP estimation.

\section{Main Result}
A key difference of ISAC problems from stand-alone estimation problems in the previous subsection is that observations are controllable by our codeword design $X^n$, i.e., if $X_i = x \in \Xc$ is chosen, the estimator's observation $Y_i$ is randomly drawn from $p(y|x, s)$. Using this nature, the Fisher information induced by transmitting $x^n$, denoted by $J_{\D, x^n}(s)$, can be simplified as follows.
\begin{align*}
	J_{\D, x^n}(s) &:= \E\left[ - \frac{\partial^2}{\partial s^2} \log p(Y^n | x^n, s) \right] \\
	&\stackrel{(a)}{=} \E\left[ - \frac{\partial^2}{\partial s^2} \sum_{i=1}^n \log p(Y_i | x_i, s) \right] \\
	&\stackrel{(b)}{=} \sum_{i=1}^n \E\left[ - \frac{\partial^2}{\partial s^2} \log p(Y_i | x_i, s) \right] \\
	&\stackrel{(c)}{=} n\sum_{x \in \Xc} \frac{n_x}{n} \E\left[ - \frac{\partial^2}{\partial s^2} \log p(Y | x, s) \right] \\
	&=: n\sum_{x \in \Xc} \frac{n_x}{n} \cdot J_{\D, x}(s) = n \E_X[ J_{\D, X}(s) ],
\end{align*}
where (a) follows since the sensing channel is memoryless, (b) follows since the order of expectation, differentiation, and sum are exchangeable, and (c) follows if $n_x$ is defined as empirical frequency of $x \in \Xc$. In the last equality, $J_{\D, x}(s)$ is defined to be the Fisher information for $s$ induced by a single observation sent from $X=x$, and the expectation is over the empirical distribution of $x^n$. Therefore, for a given $x^n$, the Fisher information by $n$-letter codewords simply reduces to the $p_X$-weighted sum of individual Fisher information terms. Then, we have the following theorem that characterizes the optimal tradeoff between information rate and (normalized) MSE. Note that $\alpha$ in the theorem directly addresses the MSE, rather than considering its proxy such as (B)CRB.

\begin{theorem} \label{thm:ATBCRB_ISAC}
	The rate-MSE decay region $\Rc$ is the set of pairs $(R, \alpha)$ such that for some $p_X$,
	\begin{align}
		R &\le \min_{s \in \Sc} I(X;\Yt), \nn \\
		\alpha &\ge \E_S \left[ \E_X^{-1} \left[ J_{D,X}(S) \right] \right]. \label{eq:ATBCRB_ISAC_bound}
	\end{align}
\end{theorem}
Before discussing the proof, it should be remarked that the inner and outer expectations are over $X \sim p_X(x)$ and $S \sim p_S(s)$, respectively. Also, the above theorem implies that the MSE decays at $\Theta(n^{-1})$ even for ISAC estimation scenarios, which is partially anticipated by a stand-alone estimation example with i.i.d.~additive Gaussian noise, e.g.,~\cite{VanTrees1968}. The comparison to the performance when na\"{i}vely applying the BCRB to our model, e.g., \cite{Xiong--Liu--Cui--Yuan--Han--Caire2023}, will be given in Cor.~\ref{cor:BCRB_outer} after the proof.

\begin{IEEEproof}[Proof of Thm.~\ref{thm:ATBCRB_ISAC}]
	\textbf{Achievability: } Consider a constant composition code (CCC) of composition $p_X$, originated from~\cite{Csiszar--Korner2011}. By our assumption on $I(X;\Yt)$ for $s$, it can be approximated within an arbitrary accuracy by fine quantization of $S$; then the data rate analysis is the same as that for compound channels. That is, by fixing the input distribution $p_X$, the CCC with composition $p_X$ achieves the data rate
	\begin{align*}
		R < \min_{s \in \Sc} I(X;\Yt) = \min_{s \in \Sc} I(p_X, p_{\Yt|X,s}).
	\end{align*}
	Due to space limitation, the detailed proof for the data rate is omitted. See~\cite{Csiszar--Korner2011} for  details.
	
	For MSE decay, note that \eqref{eq:ATBCRB} is asymptotically tight when $n \to \infty$. In this regime, 
	\begin{align*}
		J_{\DP, x^n}(s) &= J_{\D, x^n}(s) + L_{\P}(s) = n \E_X[ J_{\D, X}(s) ] + o(n),
	\end{align*}
	since the composition of $x^n$ is $p_X$ and $L_{\P}(s)$ is only determined by the prior distribution. Using it, terms in the numerator and the denominator in \eqref{eq:ATBCRB} can be simplified in the asymptotic regime as follows.
	\begin{align*}
		\E^2[ J_{\DP, x^n}^{-1}(S)] = \frac{1}{n^2} \left(\E_{S}^{2} [ \E_X^{-1}[ J_{\D, X}(s)] ]  + o(1) \right),
	\end{align*}
	\begin{align*}
		&\E[ J_{\DP}^{-1}(S)] + \E\left[ \left( \frac{\partial }{\partial S} J_{\DP}^{-1}(S) \right)^2 \right] + \E\left[ \frac{\partial^2 }{\partial S^2} J_{\DP}^{-2}(S) \right] \\
		&= \frac{1}{n} \left( \E_{S} [ \E_X^{-1} [ J_{\D, X}(s)] ] + o(1) \right) + \frac{\text{const}}{n^2} + \frac{\text{const}}{n^2},
	\end{align*}
	since $\frac{\partial }{\partial S}$ and $\frac{\partial^2 }{\partial S^2}$ only change $S$-related terms. It in turn implies that by using the ML or MAP estimators, as $n \to \infty$,
	\begin{align*}
		n \E [ (S-\sh(Y^n|x^n))^2 ] \to \E_{S} [ \E_X^{-1} [ J_{\D, X}(s)] ].
	\end{align*}
	Then, the MSE decay $\alpha$ is 
	\begin{align*}
		\alpha \le \E_{S} [ \E_X^{-1} [ J_{\D, X}(s)] ].
	\end{align*} 
	
	\textbf{Converse:} Suppose that a rate-MSE decay pair $(R, \alpha)$ is achievable by a codebook $\Cc^{(n)}$, where $n$ is possibly very large. That is, for any $\eps > 0$, there exists a length-$n$ codebook $\Cc^{(n)}$ such that $\frac{\log |\Cc^{(n)}|}{n} \ge R-\eps$, $P_e^{(n)} \le \epsilon$, and
	\begin{align*}
		\max_{x^n \in \Cc^{(n)}} n \MSE(\sh|x^n) \le \alpha + \eps.
	\end{align*}
	
	Note that there are exponentially many codewords in $\Cc^{(n)}$, while the number of types of length $n$ is at most polynomially many in $n$ \cite{Csiszar--Korner2011}. It implies that one can take a nonempty set of types $\Tc$ such that for all $p_X \in \Tc$, a subcodebook $\Cc_{p_X}^{(n)} := \{ x^n(m): \pi(x^n(m)) = p_X \} \subset \Cc^{(n)}$ satisfies
	\begin{align*}
		\frac{ \log |\Cc_{p_X}^{(n)}| }{n} \ge  R - \eps - \delta_n
	\end{align*}
	for some $\delta_n > 0$ that tends to zero as $n\to\infty$ where $\pi(x^n)$ is the type of $x^n$. In the sequel, we restrict our attention to $\Cc_{p_X}^{(n)}$.
	
	To show the coding rate converse, let $M'$ be a message taking values uniformly random from the codewords in $\Cc_{p_X}^{(n)}$, and define a chosen codeword by $\Xt^n(m)$. Then, $\Xt^n(m) \sim \tilde{p}^n(\xt^n)$ and $\Xt_i\sim \tilde{p}(\xt)$ where
	\begin{align*}
		\tilde{p}^n(x^n) &:= \frac{\mathbf{1}\{x^n \in \Cc_{p_X}^{(n)} \} } {|\Cc_{p_X}^{(n)}|} \\
		\tilde{p}_i(x) &:= \frac{1}{|\Cc_{p_X}^{(n)}|} \sum_{x^n \in \Cc_{p_X}^{(n)}} \mathbf{1}\{x_i = x\}.
	\end{align*}
	That is, $\tilde{p}^n(\xt^n)$ is the uniform distribution over $\Cc_{p_X}^{(n)}$, and $\tilde{p}_i(\xt_i)$ is the $i$-th marginal distribution of symbols on $\Cc_{p_X}$. 
	Also, let $\bar{p}(x)$ be the empirical distribution over the entire codebook $\Cc_{p_X}^{(n)}$ given by
	\begin{align*}
		\bar{p}(x) &:= \frac{1}{n} \sum_{i=1}^n \tilde{p}_i(x) = p_X(x),
	\end{align*}
	where the last equality holds since all codewords in $\Cc_{p_X}^{(n)}$ are of type $p_X$. 
	
	By applying Fano's inequality and standard converse steps to the subcodebook $\Cc_{p_X}^{(n)}$, for every $s$,
	\begin{align*}
		n(R-\eps - \delta_n) &\le \log |\Cc_{p_X}^{(n)}| \le I(M'; \Yt^n) + n \eps_n \\
		&\le n I( \bar{p}_X, p_{\Yt|X,s}) + n\eps_n \\
		&= n I( p_X, p_{\Yt|X,s}) + n\eps_n.
	\end{align*}
	The bound should hold for any $s$, which gives the coding rate 
	\begin{align*}
		R \le \min_{s \in \Sc} I(p_X, p_{\Yt|X,s}) + \eps + \delta_n + \eps_n.   
	\end{align*}
	
	For estimation converse, the ATBCRB in \eqref{eq:ATBCRB} provides a valid lower bound for any estimator $\sh$. Evaluating the lower bound for a codeword in $\Cc_{p_X}^{(n)}$ as in the achievability,
	\begin{align*}
		\MSE(\sh|x^n) \ge \frac{ \E_{S} [ \E_X^{-1} [ J_{\D, X}(s)] ] + o(1) }{n}.
	\end{align*}
	This in turn implies that
	\begin{align*}
		\alpha + \eps &\ge \max_{x^n \in \Cc^{(n)}} n \MSE(\sh|x^n) \\
		&\ge \max_{x^n \in \Cc_{p_X}^{(n)}} n \MSE(\sh|x^n) \\
		&\ge \E_{S} [ \E_X^{-1} [ J_{\D, X}(s)] ] + o(1).
	\end{align*}
	As $\epsilon$ is arbitrary, it concludes the converse proof.
\end{IEEEproof}

Note that if the BCRB in \eqref{eq:BCRB} is used in converse instead of the ATBCRB, it gives a provably strictly looser bound than Thm.~\ref{thm:ATBCRB_ISAC}, in general. To see this, applying the BCRB for converse when $n \to \infty$, it will give us the estimation performance lower bound
\begin{align}
	\alpha \ge \E_S^{-1} \left[ \E_X \left[ J_{\D,X}(S) \right] \right]. \label{eq:BCRB_ISAC_bound}
\end{align}
Then, using Jensen's inequality for $(\cdot)^{-1}$,
\begin{align*}
	\E_S^{-1} \left[ \E_X \left[ J_{\D,X}(S) \right] \right] \le \E_S \left[ \E_X^{-1} \left[ J_{\D,X}(S) \right] \right],
\end{align*}
i.e., $\eqref{eq:BCRB_ISAC_bound}$ is looser than \eqref{eq:ATBCRB_ISAC_bound}. Noting that $\Rc$ in Thm.~\ref{thm:ATBCRB_ISAC} is the optimal tradeoff, the following formal statement can be made.
\begin{corollary} \label{cor:BCRB_outer}
	Let $\Rc_{\textsf{out}}$ be the set of pairs such that
	\begin{align*}
		R &\le \min_{s \in \Sc} I(X;Y), \nn \\
		\alpha &\ge \E_S^{-1} \left[ \E_X \left[ J_{\D,X}(S) \right] \right],
	\end{align*}
	for some $p_X$. Then, $\Rc \subseteq \Rc_{\textsf{out}}$, i.e., $\Rc_{\textsf{out}}$ is an outer bound. Furthermore, $\Rc \subsetneq \Rc_{\textsf{out}}$ unless $\E_X \left[ J_{\D,X}(s) \right]$ is independent of $s$.
\end{corollary}

\section{Numerical Evaluation}
In this section, we present an illustration of Thm.~\ref{thm:ATBCRB_ISAC} using a simple example of spectrum sensing. Suppose that there are two frequency bands on which a transmitter can send an input symbol $X \in \{\pm\sqrt{P}\}$ for BPSK, $X \in \left\{\pm 3\sqrt{\tfrac{P}{5}}, \pm\sqrt{\tfrac{P}{5}} \right\}$ for 4-PAM, where $P$ is the transmission power. The first band is licensed, exclusively dedicated to the transmitter. The second band is unlicensed, where there is additive Gaussian interference with random variance drawn from a known beta distribution. Two channels are both additive white Gaussian noise (AWGN) channels. The goal is to transmit as much data as possible while accurately estimating the level of interference. Specifically, the communication and sensing channel outputs are respectively
\begin{align*}
	\Yt_i &= x_i + \Zt_i  ~~~~~~~~~ \text{if the first band is selected,}\\
	\Yt_i &= x_i + V_i + \Zt_i ~~~ \text{if the second band is selected,}
\end{align*}
and
\begin{align*}
	Y_i &= x_i + Z_i ~~~~~~~~~ \text{if the first band is selected,} \\
	Y_i &= x_i + V_i + Z_i ~~~ \text{if the second band is selected,}
\end{align*}
where $\Zt_i, Z_i \sim \Nc(0, \sigma^2)$ are the independent AWGN, and $V_i \sim \Nc(0, S)$ with $S \in [0,1] \sim \beta(a,b)$ is an interference signal. Here, $\beta(a,b)$ is the beta distribution with parameters $(a,b)$: Letting $\Gamma(\cdot)$ be the Gamma function,
\begin{align*}
	p_S(s;a,b) = \frac{\Gamma(a+b)}{\Gamma(a) \Gamma(b)} s^{a-1} (1-s)^{b-1}.
\end{align*}
In this example, we assume $a=b$ for simplicity and $a=b>2$ to meet the regularity conditions. At each symbol time, the transmitter selects one of the channels and sends a binary symbol over the chosen channel.

Since the first channel is free of interference, it is optimal for communication to send all bits directly through the first one. However, doing so prevents the estimator from estimating the interference level, as no useful sensing data is received from the second channel. On the other hand, sending bits only through the second channel is ideal for estimating the interference level but strictly degrades the communication performance. Therefore, the selection ratio between the channels must be carefully designed.

Let $t \in (0,1)$ be the fraction of the usage of the first channel. Note that a transmission symbol does not make any difference for estimating since it is given to the estimator. Hence, for each selection of channel, we assume inputs are uniformly distributed for simplicity. Note that the uniform input maximizes the information rate for BPSK. Let $C_1, C_2(s)$ respectively be the channel capacities of two channels in bits. Using these, the overall information rate at ratio $t$ is given by \cite[p.~236]{CoverT2006}
\begin{align*}
	R &\le \min_{s} I(X;\Yt) = H_2(t) + t C_1 + (1-t) \min_s C_2(s) \\
	&\stackrel{(a)}{=} H_2(t) + t C_1 + (1-t) C_2(1) \\
	&= H_2(t) + t \left(h(\Yt') - \frac{1}{2}\log (2\pi e \sigma^2) \right) \\
	&~~~~~~~~ ~~~~+ (1-t) \left( h(\Yt'') - \frac{1}{2} \log_2 (2\pi e( 1+\sigma^2 )) \right),
\end{align*}
where $H_2(\cdot)$ is the binary entropy function, $\Yt'$ and $\Yt''$ are Gaussian mixture distributions induced by input with component variance $\sigma^2$ and $1+\sigma^2$, respectively. The step (a) follows since the channel capacity for the second subchannel is minimized when the interference is maximized, i.e., $s=1$.

For estimation, once the first channel is chosen, the observation is independent of interference, i.e.,  $J_{\D,x}(s) =0$, while $J_{\D,x}(s) = \frac{1}{2(s+\sigma^2)^2}$ for the second channel. Hence,
\begin{align*}
	\E_X [ J_{\D, X}(s) ] &= \frac{t}{2(s+\sigma^2)^2}, \\
	\E_S \left[ \E_X^{-1} \left[ J_{D,X}(S) \right] \right] &=\frac{2}{t}\left( \frac{a+1}{2(2a+1)} + \sigma^2 + \sigma^4 \right).
\end{align*}

\begin{figure}[t]
	\begin{center}
		\includegraphics[width=2.5in]{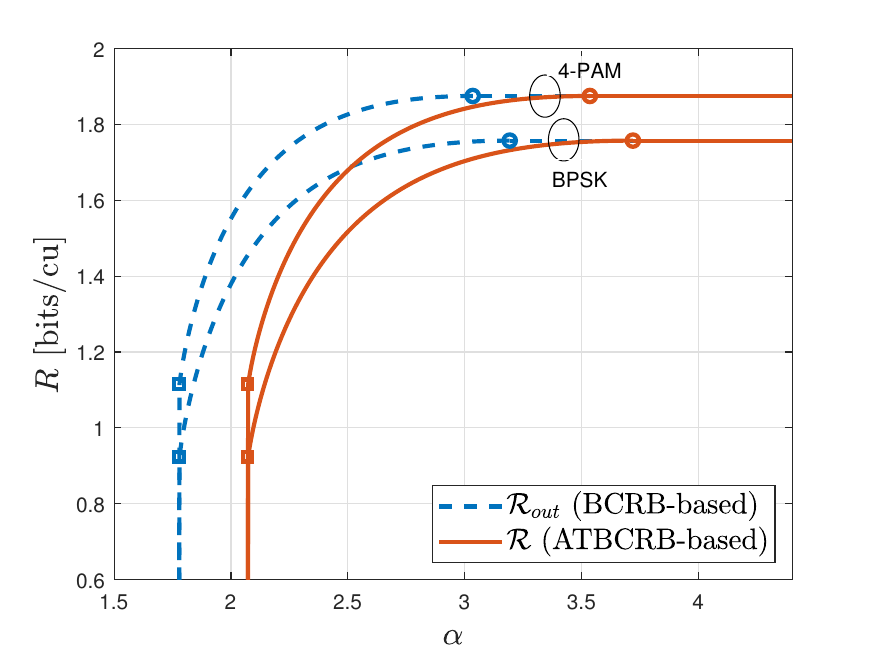}
	\end{center}
	\vspace{-0.2in}
	\caption{The optimal tradeoff region $\Rc$ in Thm.~\ref{thm:ATBCRB_ISAC} and the outer bound $\Rc_{\textsf{out}}$ using the BCRB in Cor.~\ref{cor:BCRB_outer} are plotted together. Note that $\Rc_{\textsf{out}}$ is strictly far from the optimal tradeoff, i.e., not achievable.}
	\vspace{-0.15in}
	\label{fig:rate_region}
\end{figure}

Figure \ref{fig:rate_region} shows the evaluation of $\Rc$ in Thm.~\ref{thm:ATBCRB_ISAC} (solid) and $\Rc_{\textsf{out}}$ in Cor.~\ref{cor:BCRB_outer} (dashed) for the example with $a=b=3$, $P=2$, and $\sigma^2=0.5$. Quantities that do not have analytic expressions are computed via numerical integration. Region $\Rc$ is based on the ATBCRB and is the optimal rate-MSE decay tradeoff as proven in Thm.~\ref{thm:ATBCRB_ISAC}, while $\Rc_{\textsf{out}}$ is based on the BCRB and is significantly suboptimal as expected in Cor.~\ref{cor:BCRB_outer}. Two key operating points are highlighted with circles and squares. The circles (i.e., communication-optimal points) indicate the best estimation performance achievable while maintaining the channel capacity. In contrast, the squares (i.e., estimation-optimal points) represent the maximum data rate that can be attained while maintaining the stand-alone optimal MSE.

\section{Conclusion}
This work addresses an ISAC problem where the goal of sensing is to estimate a random state that remains constant during transmission. Unlike existing literature that adopts the BCRB as a proxy for MSE, we directly characterize the optimal tradeoff between data rate and \textit{exact} MSE. Note that our characterization is optimal, whereas the BCRB-based approach is strictly suboptimal for general models. Consequently, the code design based on Thm.~\ref{thm:ATBCRB_ISAC} provides significantly better guidance. Numerical results demonstrate that the BCRB-based ISAC strategy, commonly used in the literature, might be significantly suboptimal for general Bayesian model. To the best of our knowledge, this is the first information-theoretic study that directly handles the tradeoff between information rate and \textit{exact} MSE for ISAC systems.

\end{document}